\documentclass{PoS}
\usepackage{here}

\title{Recent results on heavy quark quenching in ultrarelativistic heavy ion collisions}

\ShortTitle{Recent results on heavy quark quenching in URHIC}

\author{\speaker{P.B. Gossiaux}\\
        SUBATECH, Universit\'e de Nantes, EMN, IN2P3/CNRS, 4 rue Alfred Kastler, 44307 Nantes cedex 3, France\\
        E-mail: \email{gossiaux@subatech.in2p3.fr}}

\author{J. Aichelin\\SUBATECH}
\author{M. Bluhm\\SUBATECH}
\author{T. Gousset\\SUBATECH}
\author{M. Nahrgang\\SUBATECH \&
 Frankfurt Institute for Advanced Studies (FIAS), Ruth-Moufang-Str. 1, 60438 Frankfurt am Main, Germany}
\author{S. Vogel\\SUBATECH}
\author{K. Werner\\SUBATECH}

\abstract{In this contribution, we present some predictions for the production of D and B mesons in ultrarelativistic 
heavy ion collisions at RHIC and LHC energies and confront them with experimental results obtained so far by the STAR, PHENIX, 
ALICE and CMS collaborations. 
We next discuss some preliminary results obtained with an improved description of the medium based on EPOS initial conditions, 
and its possible implications on the nuclear modification factor and on the elliptic flow of heavy quarks.}

\FullConference{Sixth International Conference on Quarks and Nuclear Physics,\\
		April 16-20, 2012\\
		Ecole Polytechnique, Palaiseau, Paris}

\begin{document}

\section{(Short) Introduction}
Since the discovery that heavy quarks (HQ) manifest a large coupling with the hot medium created 
in heavy ions ultrarelativistic collisions, many models and theoretical 
schemes (for a review see e.g. \cite{peignesmilga, Gossiaux:2011ISMD}) have been proposed 
in order to achieve a quantitative understanding of experimental data \cite{PhenixandStar}.  
While a consensus has not been achieved yet on the most appropriate way to treat the coupling
of heavy quarks with this medium, new results from LHC experiments provide more constraints on the
various approaches and enrich the debate. In this contribution, we recall some of our results for
RHIC experiments and provide the ensuing predictions for LHC energies, both obtained
by taking the fluid dynamic approach of Kolb and Heinz \cite{KolbHeinz} (KH) to model the background
medium. We then present some preliminary study of the consequences of taking an improved 
description of the medium by choosing the EPOS initial conditions.

\section{Results at RHIC}
\begin{figure}[H] 
\centering
\includegraphics[width=6 cm]{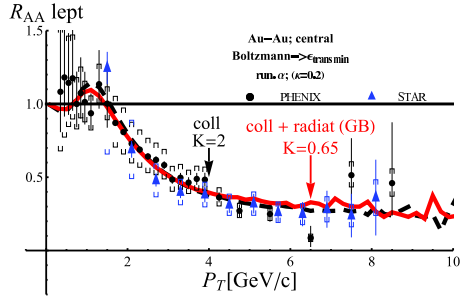}
\caption{$R_{AA}$ of non photonic single electrons for central Au-Au collisions
at RHIC energy and mid-rapidity; see \cite{Gossiaux:SQM09} 
for details.}
\label{rad_vs_col}
\end{figure}

In fig. \ref{rad_vs_col}, we recall predictions from our ``Monte Carlo at Heavy Quarks'' generator (${\rm MC}@_s{\rm HQ}$) 
for the nuclear modification
factor of non-photonic single-electrons (NPSE) stemming from D and B decay in Au-Au collisions
at $\sqrt{s}=200~{\rm GeV/c}$ using either collisional energy loss
or collisional + radiative energy loss for HQ as described in \cite{Gossiaux:SQM09}.
Both models are able to cope with the NPSE $R_{AA}$ data measured by RHIC experiments, 
at the price of a global rescaling of the interaction rate ($K=2$ and 0.65 respectively). 
As explained in \cite{Gossiaux:2011ISMD}, this model ``fragility''  has its origin in the rather stiff initial 
$p_t$-distribution of heavy quarks, which makes a sequence of many  small losses more probable than a single process involving large energy 
loss. 

One thus recovers Langevin's picture, so that the full dynamical 
evolution is encompassed in a single transport coefficient, e.g. the drag or the diffusion 
coefficient. Keeping $K$ fixed and considering other centralities or the elliptic flow
observable, one obtains a very good agreement as well, which seems to confirm this interpretation.  
In \cite{Gossiaux:2011ISMD}, we have presented the drag coefficient $A$ for the various rescaled models 
that allow to match RHIC NPSE data and have extracted values for the spacial diffusion coefficient $D_s$ of 
$2\pi T D_s\approx 1.9\pm 0.5$ at $T=200~{\rm MeV}$ and of
$2\pi T D_s\approx 2.6\pm 0.65$ at $T=300~{\rm MeV}$, for both $c$ and $b$ quarks,
in quite good agreement with recent lattice calculations\cite{Ding:2011}. 

\begin{figure}[H]
\centering
\includegraphics[width=6.5 cm]{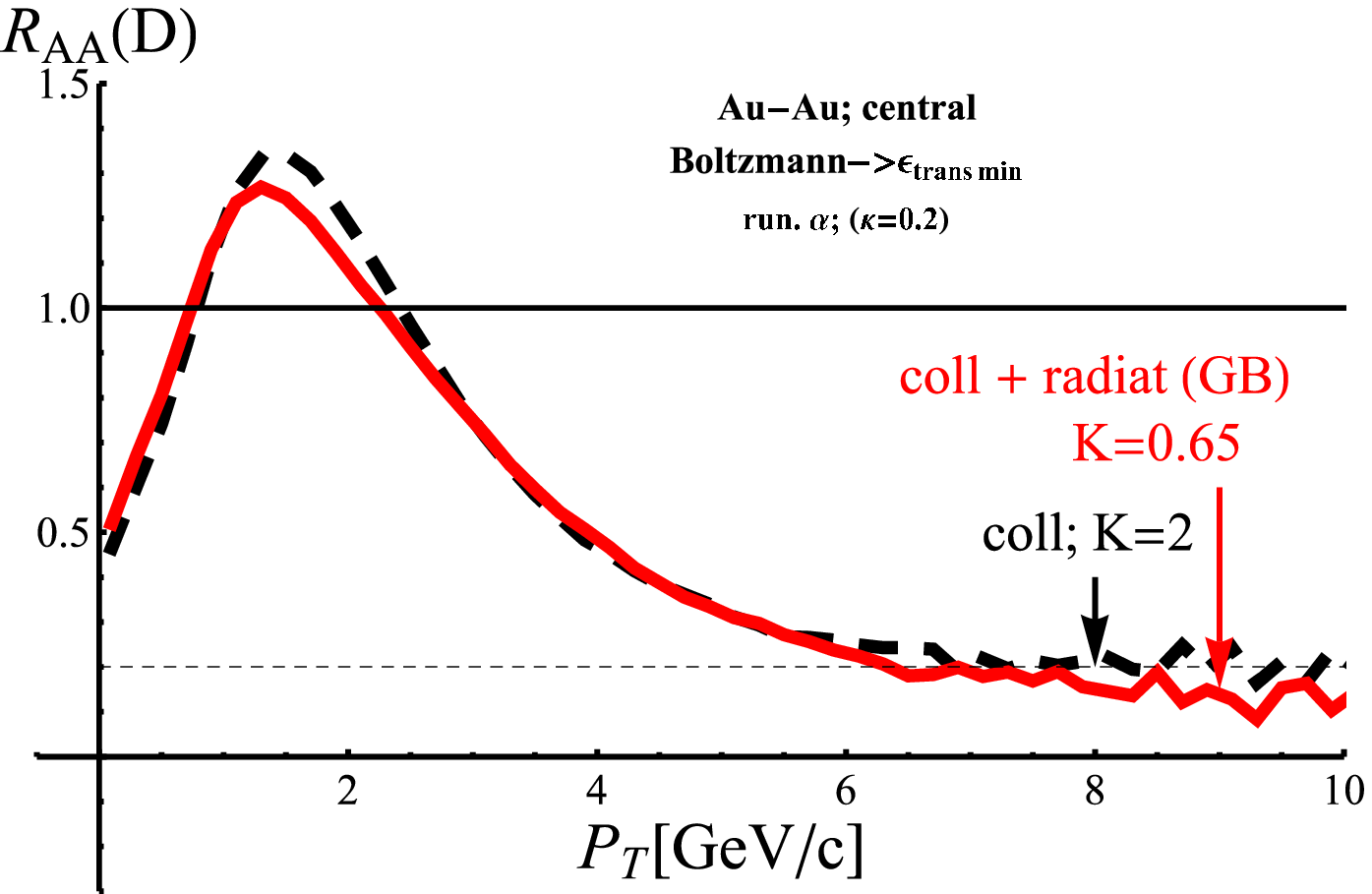}
\hspace{1cm}
\includegraphics[width=6.5 cm]{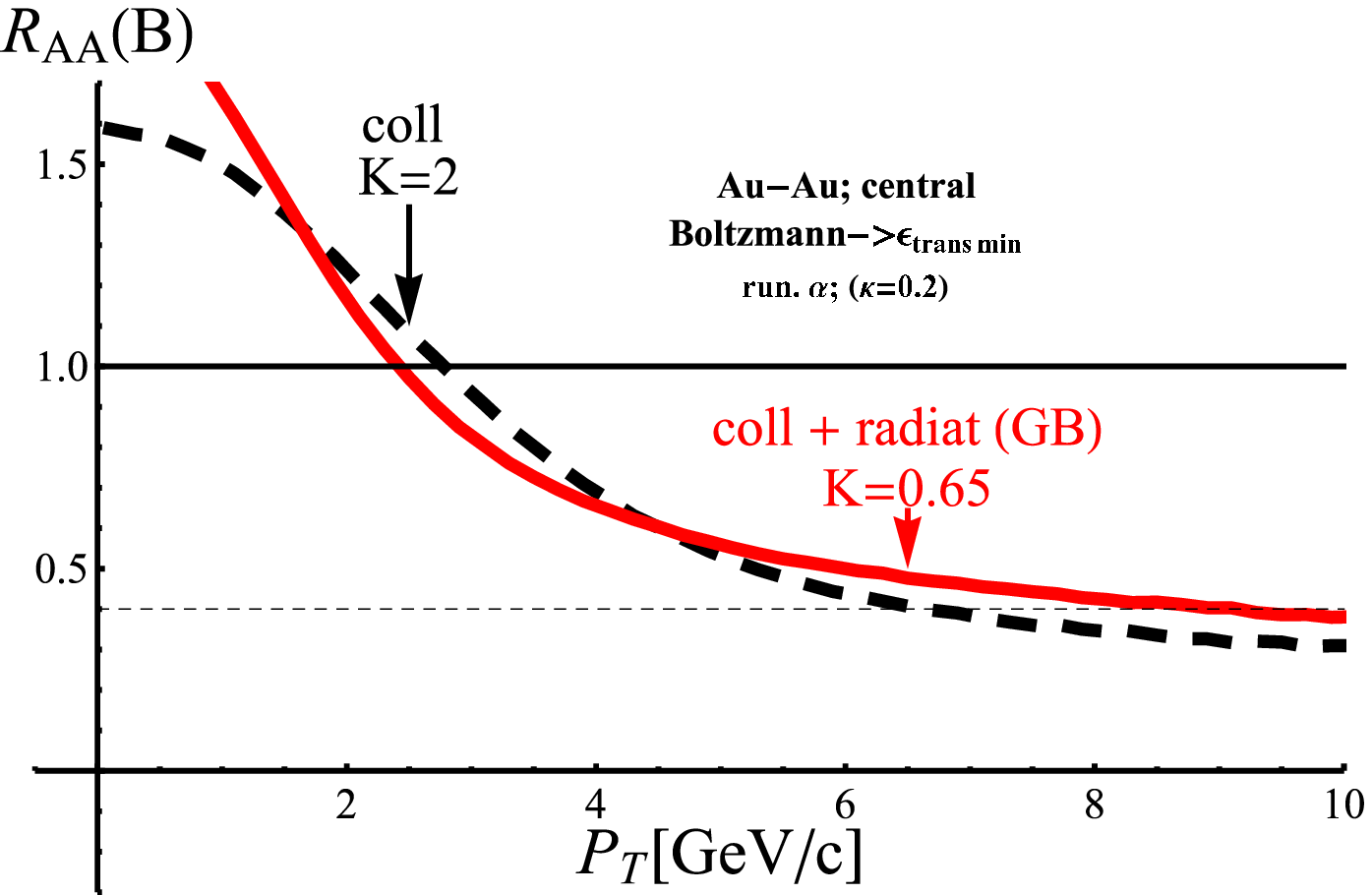}
\caption{Prediction for the $R_{AA}$ of heavy mesons according to the same models as those used in fig. 
\protect\ref{rad_vs_col}.}
\label{rad_vs_col_HM}
\end{figure}

In fig. \ref{rad_vs_col_HM}, we show our predictions separately for the D and the B mesons for
the same system as in fig. \ref{rad_vs_col}. Accordingly, one sees that it might be difficult
to distinguish between both types of energy loss mechanisms, unless future upgrades of
RHIC experiments permit to access heavy mesons with $p_t\gtrsim 10~{\rm GeV/c}$. 
We have checked that the peak observed at $p_t\approx 1.5~{\rm GeV/c}$ in $R_{AA}(D)$ is essentially
due to the radial flow of light-quarks that recombine with heavy ones through coalescence in order to 
form heavy mesons and not due to the radial flow of heavy quarks themselves (the $R_{AA}(c)$ is peaked
at $p_t=0~{\rm GeV/c}$), which explains why such peak is not observed in $R_{AA}(B)$.

\section{Predictions at LHC}
\begin{figure}[H]
\centering
\includegraphics[width=6.5 cm]{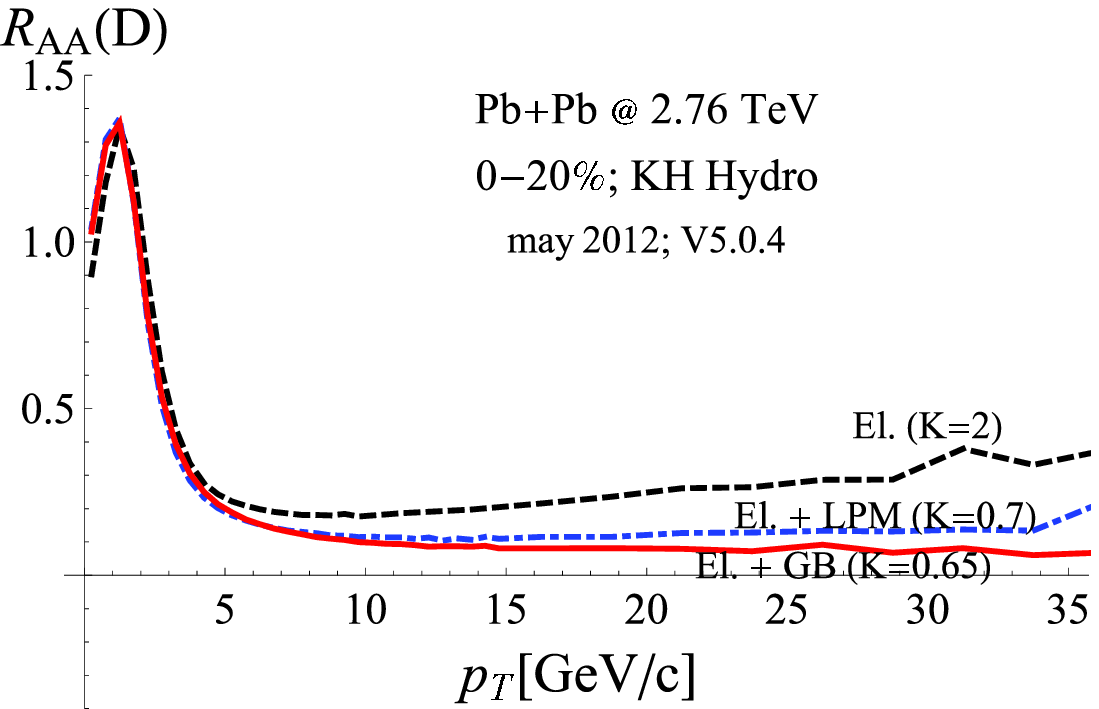}
\hspace{1cm}
\includegraphics[width=6.5 cm]{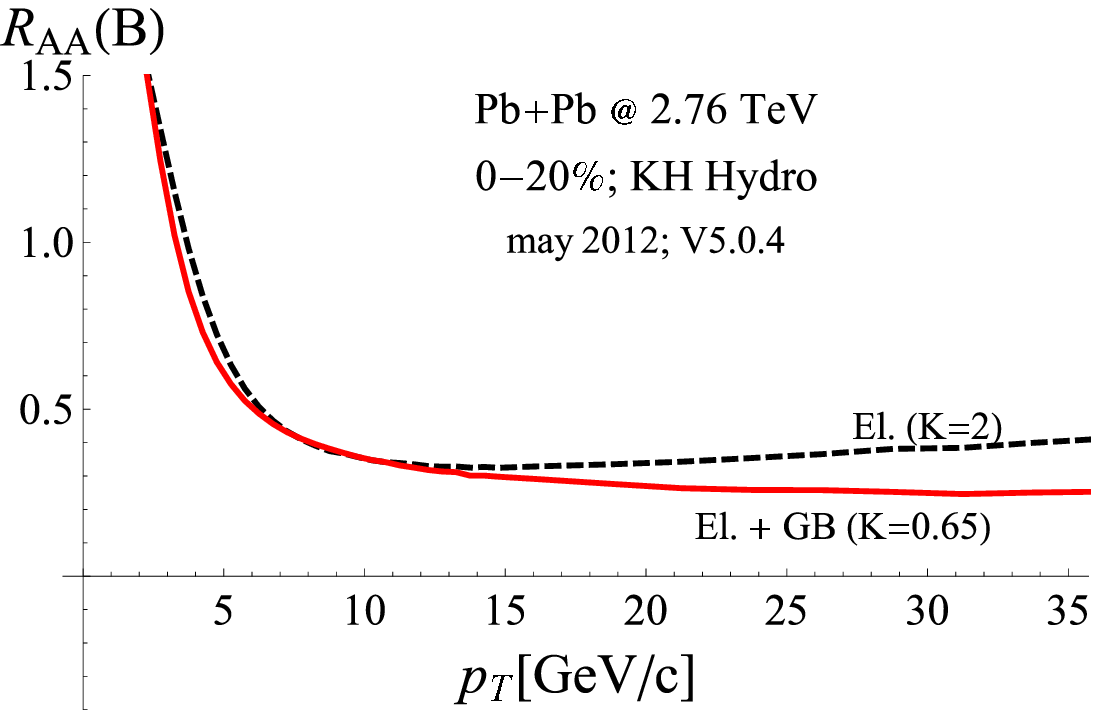}
\caption{Same as fig. \protect\ref{rad_vs_col_HM} for $\sqrt{s}=2.76~{\rm TeV}$; see text for details.}
\label{rad_vs_col_HM_LHC}
\end{figure}

We have made predictions \cite{Gossiaux:2011ISMD,PRC79, SQM2011} for the production of heavy mesons 
in Pb+Pb collisions at $\sqrt{s}=2.76~{\rm TeV}$ using the same elementary cross sections as 
for the RHIC case and adapting the initial entropy density $s_0$ in the fluid dynamics to obtain 
$\frac{dN_{\rm ch}}{dy}=1600$ at the chemical freeze out as well as the initial $p_t$-distributions 
(taken from FONLL 1.3.2).
Fig. \ref{rad_vs_col_HM_LHC} shows our latest calculations for the $R_{AA}$ of heavy mesons 
at mid-rapidity (recent predictions for the $v_2$ and the centrality dependence can be found in
\cite{Gossiaux:2011ISMD,SQM2011}). The ``El. + LPM'' curve on the left panel corresponds to our model where 
coherence effects are taken into account for the radiative energy loss by considering a single effective 
scattering center; for B mesons, those effects have little consequences.
The quenching of $B$ mesons has not been measured directly yet but can be
{\em bona fide} assimilated to the quenching of non-prompt $J/\psi$ measured
by CMS \cite{Dahms:2011}, with $R_{AA}\approx 0.35$ for $6~{\rm GeV/c}<p_t^{J/\psi}<30~{\rm GeV/c}$,
in good agreement with our predictions. For D mesons, our results show an excess of 
quenching as compared to the recent ALICE data \cite{Alice:RAAHQ} -- especially when one includes
a significant proportion of radiative energy loss -- which is under present study. 
Interesting though is the possibility to 
discriminate between collisional and radiative energy loss provided one acquires data 
in the $p_t$ range of 10-30 GeV/c.

\section{Influence of an improved medium description}
\noindent
In the following we will discuss the influence of an improved description of the fluid dynamic background which 
we choose from EPOS initial conditions \cite{Werner:2010aa}. Including a final hadronic cascade this model is able to simultaneously describe a variety of soft observables such as particle yields, spectra, flow coefficients and dihadron correlations at RHIC and LHC energies. Having the soft sector under control gives us confidence that we can reliably investigate the medium-induced energy loss of heavy quarks.
The features which distinguish the fluid dynamic evolution of the EPOS initial conditions from the previously used KH fluid dynamic background are explained in the following.
The initial conditions are obtained from a flux tube approach to multiple scattering. Each elementary scattering process is described by a parton ladder, whose final state is a longitudinal color field. The dynamics of this flux tube is described by a relativistic string. In elementary collisions the string breaking by $\bar qq$ production leads to hadron formation from the individual string segments. In nucleus-nucleus collisions the density of flux tubes is large and string segments, which are slow and/or far from the surface, are transformed to fluid dynamically evolving matter. From this procedure one obtains the initial profiles for all fluid dynamic fields including the initial transverse velocities. These latter quantities turn out to be very small and can be adjusted as a parameter.
This random flux tube distribution allows for an event-by-event treatment of the fluid dynamic evolution accounting for the fluctuating spatial structure of single events. In principle we could study the heavy quark propagation in a realistic event-by-event simulation. For the gain of statistics, however, we evolve several hundred thousand MC@sHQ runs per fluid dynamic event.

The fluid dynamic evolution is governed decisively by the underlying equation of state (EOS). While the KH fluid dynamic medium is obtained from an unrealistic EOS with a strong first order phase transition based on ideal hadron and quark-gluon gases, subsequent to the EPOS initial condition the full $3+1$d fluid dynamic simulation is performed including a parametrization of the EOS from lattice QCD \cite{Borsanyi:2010cj}. It exhibits a crossover transition between partonic and hadronic degrees of freedom in a range of temperatures between $T=145-165$ MeV \cite{Borsanyi:2010bp}.
\begin{figure}[H]
\centering
\includegraphics[width=0.45\textwidth]{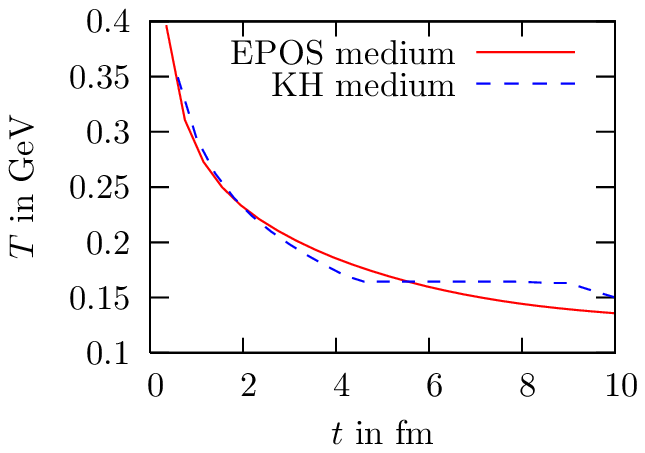}
\includegraphics[width=0.45\textwidth]{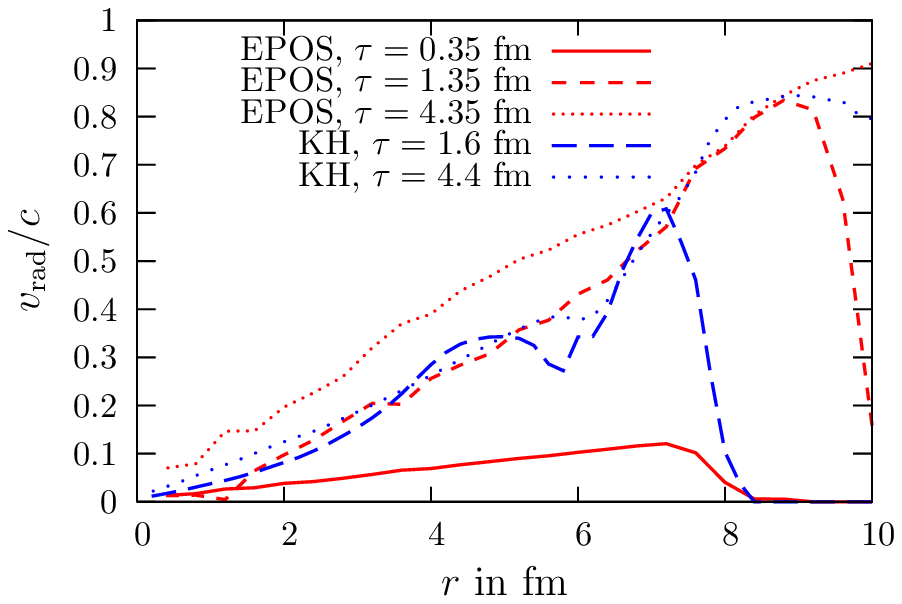}
\caption{Temperature evolution and radial velocity profiles for a central RHIC event comparing the EPOS to the KH medium.}
\label{fig:epostemp}
\end{figure}
The key ingredients for the propagation of heavy quarks are the temperature fields and the fluid velocities. These two quantities from one consistent fluid dynamic evolution are shown in fig. \ref{fig:epostemp}. The time evolution of the temperature in the center of the collision area is shown in the left plot of fig. \ref{fig:epostemp} for the fluid dynamic evolution of EPOS compared to that of KH. The initial time for the fluid dynamic expansion is $\tau_0=0.35$~fm in the case of EPOS and $\tau_0=0.6$~fm for KH. A clear difference is seen at later times, when the KH EOS exhibits a first order phase transition leading to a constant temperature of $165$~MeV, while the temperature in the crossover case keeps dropping. Although the temperature evolution above the phase transition is similar for both media, due to the different EOS the energy density, pressure and the resulting velocities are different.
 The right plot of fig. \ref{fig:epostemp} shows the radial velocity profile at different times comparing the EPOS to the KH medium. The radial velocity in EPOS is finite initially and then develops quickly toward its final value up to around $0.9c$ at the edges of the fluid. There is no initial radial velocity in the KH medium and the maximal values reached during the evolution are below those from EPOS.

\begin{figure}[H]
\centering
\includegraphics[width=0.45\textwidth]{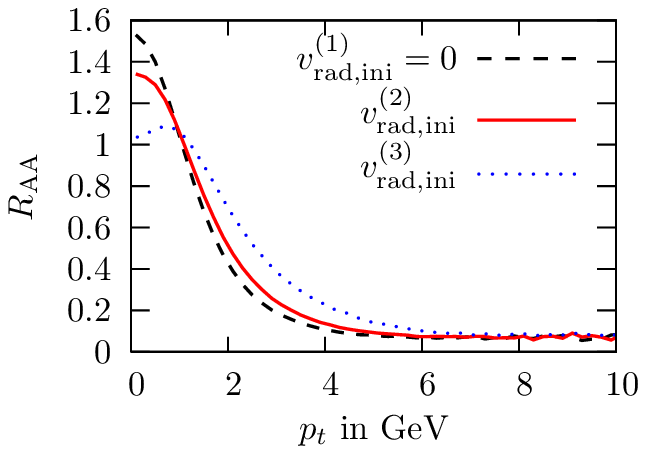}
\includegraphics[width=0.45\textwidth]{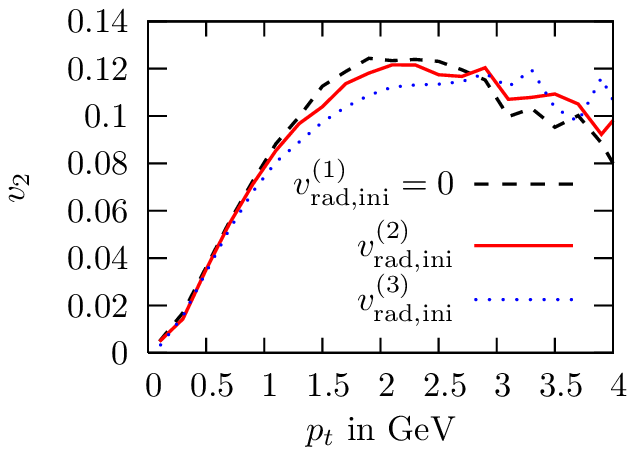}
\caption{The $R_{\rm AA}$ and the elliptic flow for charm quarks for different initial radial velocities at RHIC.}
\label{fig:raav2}
\end{figure}

We initialize the heavy quarks spatially at the nucleon-nucleon collision points given by the elementary scatterings in EPOS in the transverse plan at zero space-time rapidity.
Figure \ref{fig:raav2} demonstrates the influence of the initial radial velocity on the heavy quark observables: the nuclear modification factor $R_{\rm AA}$ (left plot) and the elliptic flow $v_2$ (right plot). These simulations are done for Au+Au collisions at a beam energy of $\sqrt{s}=200$~GeV and a centrality $0-10$\% ($R_{\rm AA}$) or a fixed impact parameter $b=7$~fm ($v_2$), 
with the ``El. + LPM'' cocktail and $K=1$. The  $R_{\rm AA}$ corresponds to the spectra of c-quarks at the transition 
temperature $T=155$~MeV, where we stop the evolution.

In the $R_{\rm AA}$ of c-quarks one can observe how the $p_t$ spectra are shifted toward higher $p_t$ for increasing initial radial velocities, where $v_{\rm rad, ini}^{(2)}$ corresponds to the standard value utilized in EPOS and for which the profiles were shown in figure  \ref{fig:epostemp}. At the largest initial radial velocity investigated, $v_{\rm rad, ini}^{(3)}$ the spectra at low $p_t$ are depleted and one finds a peak in $R_{\rm AA}$ at around $p_t\simeq 1$~GeV. For the investigated range of initial radial velocities the $R_{\rm AA}$ is influenced in the low and intermediate $p_t$ range up to $p_t\simeq7$~GeV. The quenching for high-$p_t$ particles is obviously the same irrespectively of the initial radial velocity.
A larger initial radial velocity accelerates the radial expansion of the fluid and counteracts the building up of the elliptic flow at small--intermediate $p_t$ \cite{Huovinen:2001cy}. This effect becomes more evident for particles with larger mass. Within our approach we can clearly observe this issue in the $v_2$ of c-quarks, see right plot of figure \ref{fig:raav2}. For the largest initial radial velocity we find the smallest $v_2$ up to $p_t\simeq2.5$~GeV, where $v_2$ is largest for zero initial radial velocity.

\begin{figure}[H]
\centering
\includegraphics[width=0.45\textwidth]{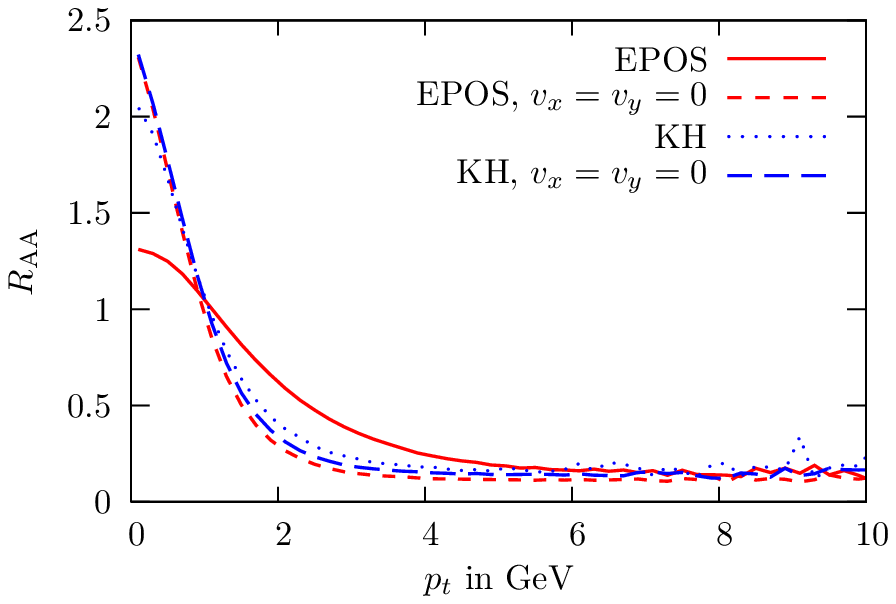}
\includegraphics[width=0.45\textwidth]{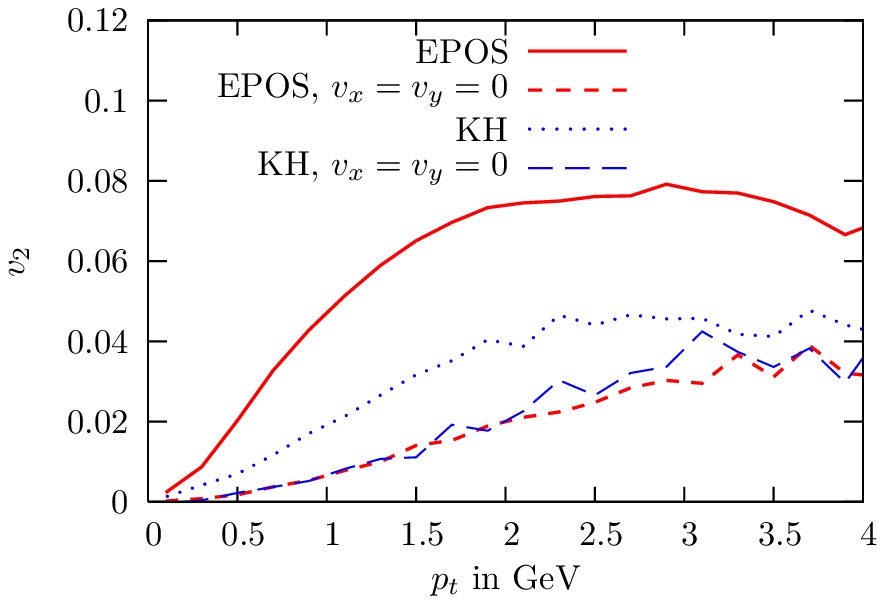}
\caption{The $R_{\rm AA}$ and the elliptic flow $v_2$ of charm quarks for the EPOS and the KH medium with and without radial velocity, central RHIC events.}
\label{fig:raav3}
\end{figure}

Finally, we compare the c-quark $R_{\rm AA}$ (10\% most central) and $v_2$ (fixed $b=7$~fm) for the EPOS and the KH medium 
in the left and right plot of figure \ref{fig:raav3} respectively. For this purpose we stop the evolution at $T=168$~MeV, 
which is above the first order phase transition in KH, and take $K=0.7$. Due to the larger radial velocity in the EPOS medium 
the spectrum of c-quarks is shifted toward higher $p_t$ compared to KH. Therefore, the c-quark $R_{\rm AA}$ at $p_t<1$~GeV is much 
lower in the EPOS medium than in the KH medium. If one switches off the radial velocity for the quark propagation by hand both 
media yield the same $R_{\rm AA}$ at small $p_t$. A slight difference remains stemming from the small differences in the 
temperature profiles, the fluctuating initial conditions and the different $\tau_0$.

If the radial velocities are switched off completely the only contribution to the elliptic flow is $p_t$ broadening over the different path lengths in $x$ and $y$ directions. Therefore, we find a reduced but finite elliptic flow $v_2$ which is of the same order in the EPOS and the KH medium. Since the 
EOS are different, the flow pattern in EPOS and KH yield significantly different elliptic flow if the full radial velocities are considered.

\section*{Acknowledgement}
The work is supported by the European Network I3-HP2 Toric, the ANR research program ``Hadrons@LHC'' (grant ANR-08-BLAN-0093-02) and the Pays de la Loire research project TOGETHER.

\end{document}